\documentstyle{article}
\begin{document}

\thispagestyle{empty}
\begin{flushright}
BONN-TH-98-23\\
hep-th/9811239
\end{flushright}

\vspace{1.5cm}

\begin{center}
{\LARGE The associative algebras of conformal field theory}
\vspace{1.0cm}

{\large D. Brungs\footnote{email brungs@avzw02.physik.uni-bonn.de}
 and W. Nahm\footnote{email werner@avzw02.physik.uni-bonn.de}} 

\vspace{0.7cm} 
{Physikalisches Institut, Universit\"at Bonn\\
 Nu\ss allee 12, 53115 Bonn, Germany}
\vspace{0.5cm}

{November 1998}
\vspace{0.5cm}
\end{center}

\begin{abstract}

Modulo the ideal generated
by the derivative fields, the normal ordered product
of holomorphic fields in two-dimensional conformal
field theory yields a commutative and associative algebra.
The zero mode algebra can be regarded as a deformation of the 
latter. Alternatively, it can be described as an associative
quotient of the algebra given by a modified normal ordered
product. We clarify the relation of these structures
to Zhu's product and Zhu's algebra of the mathematical literature.

\end{abstract}

\section{Introduction}

Apart from free field theories, conformally invariant theories in
two dimensions were the first quantum field theories for which an
elegant and efficient mathematical theory was developped (\cite{Bor1}, 
\cite{FLM}, \cite{God}, for a recent proposal on axiomatic 
foundation of conformal field theory see \cite{GG}, where Zhu's algebra is 
also discussed).
Deficiencies of communication between mathematicians and physicists,
however, led to parallel developments in both communities, which are
inconvenient and often confusing.

A glaring example is the classification of representations
of the W-algebra of holomorphic fields. Since
the early days of conformal field theory, 
physicists have used the zero mode algebra for that purpose (e.g.\ \cite{Zam}, \cite{BG}, 
\cite{KZ}).
Mathematicians talk about Zhu's algebra instead, and define
the latter with the help of Zhu's product (\cite{Zhu0}, \cite{DLM}). Its seems that it took them
a while to discover that the two algebras are naturally isomorphic,
but the fact has been stated in their work
(\cite{FZ}, remark after proposition 1.4.2). This fact has been
overlooked by physicists who started to use Zhu's algebra
under the impression that it is a new structure (e.g.\ \cite{EG}). We also failed to 
notice the remark until we rediscovered the isomorphism by ourselves.

\section{Operator product expansion and normal ordered products}

The main reason for
the unnecessary duplication of concepts seems to be the lack of 
familiarity of mathematicians with the operator product expansion 
(OPE).

Consider holomorphic conformal fields $\phi$ and $\chi$ of conformal
dimensions $h$ and $h'$. Their Fourier expansion is of the form
$$\phi(z)=\sum_n z^{n-h}\phi_n$$
and their OPE can be written as
$$\phi(z)\chi(w)=\sum_r (z-w)^r \psi^r(w),$$
where the holomorphic fields $\psi^r$ have dimensions $h+h'+r$.
In particular, the sum can be restricted to integers $r\geq -h-h'$.
The field  
$$\psi^0(w)=\oint_{C(w)} {dz\over 2\pi
i(z-w)}\phi(z)\chi(w),$$ 
$C(w)$ a circle around $w$, is the standard normal ordered product
of $\phi$ and $\chi$. One common notation is $\psi^0=N(\phi,\chi)$. Writing $C(w)$
as the difference of cycles around zero, one obtains the Fourier coefficients  
$$\psi^0_n=\sum_{m\geq h}\phi_m\chi_{n-m}+\sum_{m<
h}\chi_{n-m}\phi_m.$$ 
We grade by the energy, such that creation 
operators have positive Fourier index.

Instead of splitting the sum at $m=h$ one can use any other
fixed value of $m$. The resulting field differs from $N(\phi,\chi)$ by
a linear combination of some fields of lower dimension, namely the
$\psi^r$ with $r<0$. Splitting at $m=0$ yields   
$$N_0(\phi,\chi)_n=
\sum_{m\geq 0}\phi_m\chi_{n-m}+\sum_{m< 0}\chi_{n-m}\phi_m.$$  
The corresponding field can be obtained from its value at 1 by translation. Since
$$N_0(\phi,\chi)(1)=\sum_{n} N_0(\phi,\chi)_n 
= \oint_{C(1)} {dz\over 2\pi i(z-1)} z^h \phi(z)\chi(1),$$
one obtains

$$N_0(\phi,\chi)(0)= \oint_{C(0)} {dz\over 2\pi i z} (1 + z)^h \phi(z)\chi(0).$$
The isomorphism $\psi(0)|{\mit vac}\rangle = |\psi\rangle$ between fields and states yields

$$|N_0(\phi,\chi)\rangle = \oint_{C(0)} {dz\over 2\pi i z} (1 + z)^h \phi(z)|\chi\rangle,$$
which coincides with Zhu's definition of the product $\phi \ast \chi$.

One can modify the $N_0$-product by scaling its dimension $h+h'-r$
components by $\lambda^r$:

$$\oint_{C(0)} {dz\over 2\pi i z} (1 + \lambda z)^h \phi(z)|\chi\rangle.$$
For $\lambda=0$ one recovers $N(\phi,\chi)$. Thus the $N_0$-product can be regarded as
a deformation of the $N$-product. 

The derivative of the normal ordered product $N$ is given by

$$\partial N(\phi,\chi)= N(\partial\phi,\chi)+N(\phi,\partial\chi),$$
whereas in the case of the $N_0$-product $\partial$ has to be replaced by
$L_1 + L_0$ on either side of the equation. 

\section{Zhu's commutative algebra}

The commutator $N(\phi,\chi)-N(\chi,\phi)$ 
has Fourier coefficients which are given by the commutators 
 $[\phi_m,\chi_n]$.
The latter are spanned by Fourier components $\Psi_{m+n}$ of fields
of dimension $h(\Psi)<h+h'$. Since the commutator field has dimension
$h+h'$, it is a linear combination of derivative
fields $\partial^s\Psi$, with $s=h+h'-h(\Psi)$.
Similarly, the $N$-associator
$$A(\phi,\chi,\psi)=N(\phi,N(\chi,\psi))-N(N(\phi,\chi),\psi)$$
can be calculated in terms of commutators of the
Fourier coefficients of $\phi$, $\chi$, and $\psi$. Thus it is a linear
combination of fields of the form $N(\Phi,\partial^s\Psi)$,
with $s>0$ and $h(\Phi)+h(\Psi)=h(\phi)+h(\chi)+h(\psi)-s$. 
In other words, modulo the ideal generated by the derivative fields,
the $N$ product yields a commutative, associative algebra.
Zhu defined this algebra
in section 4.4 of \cite{Zhu0}, but didn't give it a name.
Physicists seem to have overlooked its significance. Provisionally,
we shall call it Zhu's commutative algebra. The zero mode algebra is
a deformation of it, as we shall see.

\section{Zhu's algebra and the algebra of zero modes}

We have seen that Zhu's product is just the normal ordered product $N_0$.
Its properties can be derived easily from this fact. Of
particular interest is the associator
$$A_0(\phi,\chi,\psi)=
N_0(\phi,N_0(\chi,\psi))-N_0(N_0(\phi,\chi),\psi).$$
One has

\begin{eqnarray*}
A_0(\phi,\chi,\psi)_n&=&\sum_{k<0}\sum_{l=0}^{-k-1}
[\phi_{n-k-l},\chi_l]\psi_k
-\sum_{k\geq 0}\psi_k\sum_{l=-k}^{-1}[\phi_{n-k-l},\chi_l]\\
&&+\sum_{k<0}\sum_{l=k}^{-1}[\phi_{n-l},\psi_{l-k}]\chi_k -
\sum_{k\geq 0}\chi_k\sum_{l=0}^{k-1}[\phi_{n-l},\psi_{l-k}].
\end{eqnarray*}
Since $\sum_{l=0}^{-k-1}[\phi_{m-l},\chi_l]$ and its continuation
$-\sum_{l=-k}^{-1}[\phi_{m-l},\chi_l]$ is a polynomial in $k$ which 
vanishes for
$k=0$, the first two terms on the right hand side yield normal ordered products
involving fields with Fourier components $k^s\psi_k$, $s>0$, i.e.\ 
linear combinations of normal ordered products of the fields
$\partial^s\psi+(h(\psi)+s-1)\partial^{s-1}\psi$. 
The other remaining terms on the r.h.s. yield normal ordered products
of $\partial^s\chi+(h(\chi)+s-1)\partial^{s-1}\chi$. These fields
can be written in the form $\partial\Phi+h(\Phi)\Phi$. Modulo normal
ordered  products of such
fields, the product $N_0$ is associative. This has been discovered by
Zhu, though his calculations were not very transparent. The resulting
quotient algebra has been called Zhu's algebra by mathematicians,
but we see no reason to introduce this term into physics.

Actually, Zhu's formalism was a bit different, but the equivalence is
easy to check. He calculated modulo fields of the form   
$$\oint_{C(0)} {dz\over 2\pi iz^2}\Phi(z)(1+z)^h|\Psi\rangle.$$ 
When we write $(1+z)^h/z^2=(1+z)^{h+1}/z^2-(1+z)^h/z$ and get rid of
$z^{-2}$ by a partial integration, this field is reduced to
$(\partial\Phi+h(\Phi)\Phi)\ast\Psi$.

On the ground state vectors $v$ of a representation of the OPE
one has $$N_0(\phi,\chi)_0v=\phi_0\chi_0 v.$$
This yields a homomorphism of Zhu's algebra to the zero mode algebra.
Since the latter is associative, the $N_0$ associators have to
be mapped to  zero. Indeed, $(\partial \Phi)_0 +h(\Phi)\Phi_0=0$.

Every representation of the zero mode algebra induces a lowest energy
representation of the $W$-algebra of holomorphic fields. If one admits 
infinite ground state degeneracies, one such representation is
the adjoint one, which is faithful. This means that Zhu's algebra is
naturally isomorphic to the zero mode algebra, such that we see no
need for a terminological distinction. It is not clear, however, if
one could make a similar argument by using finite dimensional
representations only, apart from the obvious case when the zero mode
algebra itself is finite dimensional. In the latter case the theory
is rational.

As a vector space, the zero mode algebra is filtered by the
conformal dimension. It is not graded, since sums of type
$\partial\Phi+h(\Phi)\Phi$ are not homogeneous. The corresponding
graded vector space is just the space of all holomorphic fields
modulo the derivative fields. The corresponding algebra is Zhu's
commutative and associative algebra discussed above. 

\section{Conclusion}

We hope that the present paper will stop the duplication
of work in this corner of conformal field theory. Apparently,
all of us should work a bit harder to get through the communication
barrier. In particular, mathematicians should strive to be less
clumsy, and we, perhaps, a bit less sloppy.

\bibliographystyle{alpha}
\bibliography{paper}

\end{document}